\def\lam{$\Lambda$}
\def\kmN{($K^-,N$)}
\def\kmn{($K^-,n$)}
\def\kmp{($K^-,p$)}

\def\c12{$^{12}$C}
\def\c12l{$^{12}_\Lambda$C}

\def\pik{$(\pi^+,K^+)$}

\def\pikli6{$^{6}{\rm Li}(\pi^+,K^+)$}

\documentstyle[preprint,eqsecnum,aps,epsf]{revtex}

\begin{document}
\preprint{HEP/123-qed}
\title{Kaonic nuclei excited by the \kmN\ reaction  }
\author{Tadafumi Kishimoto}
\address{Department, of Physics, Osaka University, Toyonaka, Osaka, 
560-0043, Japan}
\date{\today} 
\maketitle  
\begin{abstract} 
We show that kaonic nuclei can be produced by the \kmp\ and \kmn\ 
reactions.   The reactions are shown to have cross sections experimentally 
measurable.  The observation of the kaonic nuclei gives a kaon-nucleus 
potential which answers the question on the existence of kaon condensation 
in dense nuclear matter especially neutron stars. 
\end{abstract} 

\narrowtext

The kaon-nucleon interaction at low energy region is particularly important 
nowadays because of the current interrest in the dense nuclear matter in 
neutron stars where so-called kaon condensed state may be achieved by a 
strong attractive interaction \cite{k_n,brown}.  The existence of the kaon 
condensed state softens the equation of state (EOS) of nuclear matter in 
the neutron stars and reduces their calculated maximum mass above which the 
neutron stars become black holes.  The observed mass distribution of the 
neutron stars agrees with the calculation with this softening \cite{prak}.   
The introduction of strange hyperons in the EOS gives a similar softening.   
Strangeness is essential in both cases although quantitative understanding 
of the EOS requires a knowledge of both the kaon-nucleon- and hyperon-nucleon 
interactions at high density \cite{ellis}.  The kaon-nucleon interaction, in 
particular, is known quite poorly experimentally.  

Recently, effective kaon mass in dense nuclear matter has been derived by 
the Chiral SU(3) effective Lagrangian including $\bar{K}N$, $\pi \Sigma$, 
$\pi \Lambda$ systems\cite{waas}.   Such a theoretical model reproduces well 
the \lam (1405) as a $\bar{K}N$ bound state due to the strong $\bar{K}N$ 
attractive interaction \cite{waas,siegel}.  The $\bar{K}N$ interaction makes 
the $K^-$ feel a strong attractive potential in nuclei which consequently 
leads to 
the existence of deeply bound kaonic nuclei.  The \lam (1405), however, can 
also be interpreted as a three-quark state with $\ell=1$ excitation.  In this 
case no attractive $\bar{K}N$ interaction is relevant and the deeply bound 
kaonic nuclei don't necessary exist.  

Experimental data of $K^-$ optical potential mostly come from kaonic atoms.  
The shifts and widths of atomic levels affected by the strong interaction 
were reproduced by introducing an appropriate optical potential in addition 
to the Coulomb interaction.  
Recent extensive analysis of kaonic X-ray data concludes that the 
potential is strongly attractive \cite{batty-f-g}.  The derived depth is 
around -200 MeV which opens a possibility of kaon condensation at around 
three times normal nuclear density.  
Derivation of the optical potential from the kaonic atom data is, however, 
subtle since the atomic state is sensitive only to the phase shift of $K^-$ 
wave function at the nuclear surface.    The phase shift alone cannot 
determine the depth of the potential since the $K^-$ wave function has an 
ambiguity in number of nodes in the nucleus especially when the potential 
depth is quite deep.   The strong imaginary part of the potential further 
obscures the nodes.   Earlier studies with a different treatment of the 
nuclear surface gave much shallower potentials of - $80 \sim 90$ MeV 
\cite{batty-f-g} which tend to exclude the kaon condensation in the neutron 
stars.

Heavy ion reactions have been studied to derive the $K^-$ optical potential 
\cite{barth}.  Enhanced $K^-$ production in the reactions suggests a strong 
attractive interaction although quantitative argument requires understanding 
of details of the reaction mechanism \cite{li}.  

The $\bar{K}N$ interaction has been derived from kaon scattering 
experiments.  However, the available low-energy data are insufficient for 
unique multichannel analysis and the existence of \lam (1405) makes the 
extrapolation of the amplitude below the threshold complicated\cite{martin}.  
Recent theoretical calculations on the kaon interaction in nuclei predict an 
attractive interaction although they are still controversial quantitatively 
and existence of kaon condensation in neutron stars is as yet inconclusive 
\cite{ramos,lutz}.

If $\bar{K}$-nuclear potential is as attractive as derived from the kaonic 
atom stuides suggest \cite{batty-f-g}, then deeply-bound kaonic nuclei 
should exist.  The observation of kaonic nuclei gives directly the $K^-$ 
optical potential and gives decisive information on the existence of kaon 
condensation in neutron stars.  
We show the general properties of the kaonic nuclei and that the $(K^-,N)$ 
reaction can excite them with cross section experimentally measurable.

Energies and widths of kaonic nuclei are calculated with the potential given 
by the kaonic atom data.  For the analysis of mesonic atoms the Klein-Gordon 
equation is usually used \cite{batty-f-g}.   Here we use the Schr\"{o}dinger 
equation with harmonic oscillator potential.  It is a crude approximation 
although it is good enough for the present purpose.   We are interested in 
gross structure of levels and an order-of-magnitude estimate of the cross 
section for the deeply bound state.  
For the moment we take the potential depth -200 MeV given by kaonic atom.  
It is roughly four times deeper than that for nucleon and the kaon mass 
is about half of that of a nucleon.  Thus the major shell spacing 
($\hbar \omega_K$) is $\sqrt{8}$ times the 40A$^{-1/3}$ frequently used for 
nucleon.  Since the kaon has no spin, no spin dependent splitting has to be 
considered.  

The $\hbar \omega_K$ is roughly 40 MeV, for instance, for the kaonic 
$^{28}_K$Si nucleus.  The $1s$ state appears at around -140 (${3\over 2} 
\hbar \omega_K -200 $) MeV bound, which is the deepest bound state ever 
observed in nuclear physics.  If the potential shape is closer to the square-well it 
appears deeper.   In order to observe the state its width has to be 
reasonably narrow.  The width is given by the imaginary part of the 
potential, which decreases for the deeply bound state and is around 
10 MeV  \cite{waas,batty-f-g}.  The narrow width is understandable since 
dominant conversion channels like $K N \rightarrow \pi \Sigma$ or $K N 
\rightarrow \pi \Lambda$ are energetically almost closed for such a 
deeply-bound state.   Kaon absorption by two nucleons ($K N N \rightarrow 
Y N$) gives little width since two nucleons have to participate to the 
reaction.   Even though the width is twice wider the $1s$ state 
should be seen well separated since the next excited state ($1p$) is 
expected to appear 40 MeV higher.

The $(K^-,N)$ reaction where a nucleon ($N$) is either a proton or a neutron 
is shown schematically in figure 1.   The nucleon is knocked out in the 
forward direction leaving a kaon scattered backward in the vertex where the 
$K + N \rightarrow K + N$ takes place.  This reaction can thus provide a 
virtual $K^-$ or $\bar{K^0}$ beam which excites kaonic nuclei.  This feature 
is quite different from other strangeness transfer reactions like 
$(K^-, \pi)$, $(\pi^\pm ,K^+)$ and $(\gamma , K^+)$ extensively used so 
far.  They primarily produce hyperons and thus are sensitive to states 
mostly composed of a hyperon and a nucleus.   

The momentum transfer, which characterizes the reaction, is shown in figure 2.  
It depends on the binding energy of a kaon.  We are interested in states well 
bound in a nucleus ($BE = 100 \sim $150 MeV).  The momentum transfer for the 
states is fairly large ($q= 0.3 \sim 0.4$ GeV/c) and depends little on the 
incident kaon momentum for $P_K = 0.5 \sim 1.5$ GeV/c, where intense kaon 
beams are available.  Therefore one can choose the incident momentum for 
the convenience of an experiment.  It is a little beyond the Fermi momentum 
and the reaction has characteristics similar to the \pik\ reaction for 
hypernuclear production where so-called stretched states are preferentially 
excited \cite{dover}.

Recently deeply bound $\pi^-$ atoms were observed by the $(d, ^3\rm{He})$ 
reaction \cite{yamazaki}.   A small momentum transfer ($\sim 60 MeV/c$) was 
vital to excite the atomic states which were typically characterized by 
the size of the atomic orbits.   If one wishes to excite kaonic atoms, a 
momentum transfer less than 100 MeV/c is desirable.  It is achieved by 
kaon beams less than 0.4 GeV/c where available beam intensity is very 
small.   The repulsive nature of the $\pi$-nucleus interaction allows no 
nuclear state although the strong attractive $\bar{K}$-nucleus potential 
makes kaonic nuclei exist.   The $(K^- ,N)$ reaction can excite the deeply 
bound kaonic nuclei with large cross section in spite of the large momentum 
transfer of the reaction.  For the excitation of nuclear states the momentum 
transfer is typically characterized by the Fermi momentum.  

The $(K^- ,N)$ reaction on deuteron is the simplest reaction by which one 
can study the $\bar{K}N$ component of excited hyperons.  The $d(K^- ,p)$ 
reaction excites $K^- n$ states which can only have $I=1$.  On the other 
hand $d(K^-, n)$ reaction excites a $K^- p$ state which can have either 
$I=1$ or $I=0$.  Cross sections to the excited hyperons depend on their 
$\bar{K}N$ component.  For instance, the well known \lam (1405 MeV) should 
be abundantly excited by the \kmn\ reaction if it is a $\bar{K}N$ bound 
state with $I=0$ as usually believed.   The $d(K^- ,p)$ 
reaction, in particular, gives information on the $K^- n$ interaction below 
the threshold, which plays decisive role on the kaon condensation in the 
neutron stars.  

We adopt here the distorted wave impulse approximation (DWIA) to evaluate 
the cross section.   The DWIA calculation requires: (a) distorted waves for 
entrance and exit channels, (b) two body transition amplitudes for the 
elementary $(K^-, N)$ process, and (c) a form factor given by initial 
nuclear and kaonic-nuclear wave functions.   Relevant formulas for the 
calculation can be found elsewhere \cite{dover}.  

The differential cross section in the laboratory system for the formation of 
kaonic nucleus is given by 
\begin{equation}
{d \sigma \over d \Omega}= \left({d \sigma \over d \Omega}\right)^{K^-N 
\rightarrow N K^-}_{L,0^\circ } N_{eff} .
\label{eq:cross section}
\end{equation}
It is given by the two body laboratory cross section multiplied by the 
so-called effective nucleon number ($N_{eff}$).  

We first use the plane wave approximation to evaluate $N^{pw}_{eff}$.  At 0 
degrees, where only non-spin flip amplitude is relevant, $N^{pw}_{eff}$ is 
given by 
\begin{equation}
N^{pw}_{eff}= (2J+1)(2j_N+1)(2\ell_K+1) 
\left( \begin{array}{ccc} \ell_{K} & j_{N} & J \\
0 & -{1 \over 2} & {1 \over 2} \end{array} \right)
F(q) .
\end{equation}
In this equation we assumed that a nucleon in a $j_N$ orbit is knocked out 
and a kaon sits $\ell_K$ state making transition from $0^+$ closed shell 
target to a spin $J$ state.  Here the form factor $F(q)$ is given by the 
initial nucleon and final kaon wave functions as 
\begin{equation}
F(q)= \left( \int r^2 dr R_K(r) R_N(r) j_L(qr) \right)^2 , 
\end{equation}
where $L= J\pm {1 \over 2}$ is the transferred angular momentum.  

For an oscillator potential of radius parameter $b$, the radial wave 
function is 
\begin{equation}
R_\ell (r) =c_\ell (r/b)^\ell e^{-r^2/2b^2}
\label{eq:formf}
\end{equation}
for nodeless states, where $c_\ell = [2^{l+2}/b^3 \sqrt{\pi} (2l+1) 
! ! ]^{1/2}$.  In the present case it is enough to consider natural parity 
stretched states with $L= \ell_N + \ell_K$ since the transferred momentum 
$q$ is larger than the Fermi momentum.  The form factor (Eq. \ref{eq:formf}) 
is well known for the harmonic oscillator wave function \cite{dover} as 
\begin{equation}
F(q)={(2Z)^L e^{-Z} \over [(2L+1)! ! ]^2} 
{[\Gamma (L+3/2)]^2 \over \Gamma (\ell_K +3/2) \Gamma (\ell_N + 3/2)}
\end{equation}
with $Z= (b q)^2/2$, where the radius parameter $b={m \omega \over \hbar}$ has 
to be replaced by 
\begin{equation}
{2\over b^2} = {1 \over b_N^2} + {1 \over b_K^2} 
\end{equation}
to account for the different radius parameters for the nucleon ($b_N$) and 
the kaon ($b_K$) where ${1 / b_K^2} = \sqrt{8} / b_N^2$.  
$N^{pw}_{eff}$ is further reduced by the distortion of incoming and outgoing 
waves as 
\begin{equation} 
N_{eff} = N^{pw}_{eff} D_{eik} \ .
\end{equation}
The distortion $D_{eik}$ is estimated by the eikonal absorption where the 
imaginary parts of the K$^-$ and proton optical potentials are given by 
their total cross sections with nucleons.   At P$_K$= 1 GeV/c, total cross 
sections of $K^-$-nucleon and $p$-nucleon are almost the same and we take 
both to be 40 mb.  The small radius parameter $b$ indicates larger cross 
sections through the high momentum component; we thus evaluated $N_{eff}$ 
for $b_K= b_N$ also as the smallest value.

The cross section of the elementary reaction was given by the phase shift 
analysis of available data\cite{gopal}.   Here we need to consider only the 
non-spin flip amplitude ($f$) as explained above.  Since the kaon and nucleon are 
isospin ${1 \over 2}$ particles there are I=0 ($f^0$) and I=1 ($f^1$) 
amplitudes.  The amplitudes for elastic and charge exchange scattering are 
represented by appropriate linear combinations of the isospin amplitudes as 
\begin{eqnarray}
f_{K^-n \rightarrow K^- n} = f^1 \ , \\
f_{K^-p \rightarrow K^- p} = {1 \over 2}(f^1 + f^0) \ , \\
f_{K^-p \rightarrow \bar{K^0} n} = {1 \over 2}(f^1 - f^0) \ .
\end{eqnarray}
The CM differential cross section of the three reactions at 180$^\circ$ are 
shown in figure 3 as a function of incident kaon momentum.  The cross 
sections depend on the incident momentum.  For instance, the $K^-p 
\rightarrow K^- p$  reaction has a distinct peak at around 1 GeV/c.  We thus 
take 1 GeV/c for the incident kaon momentum.  Since the target nucleon is 
moving in a nucleus, Fermi averaging has to be made for the two body cross 
section which smears the fine momentum dependence.  The CM cross section is 
reduced by 20 to 30 \% depending on models for this averaging.  We take 
$\sim$10 mb/sr as the CM cross section at 1 GeV/c.  

Here we consider I=0 symmetric nuclei as targets.  The \kmp\ reaction 
produces only an I=1 state; on the other hand the \kmn\ reaction can produce 
both I=0 and 1 states.   The $\bar{K}N$ system is strongly attractive in the 
I=0 channel though not so much in the I=1 channel.   The kaon-nucleus 
potential is an average of both channels and thus depends little on the 
total isospin of kaonic nuclei.  Consequently we expect that the I=0 state 
produced by the \kmn\ reaction appears at nearly the same excitation energy.  
The elementary cross section for the \kmn\ reaction in eq 
(\ref{eq:cross section}) becomes the sum of the $K^- n \rightarrow K^- n$ 
and $K^- p \rightarrow \bar{K^0} n$ cross sections.   The incoherent sum of 
the two cross sections may not be inappropriate for the evaluation since 
the $K^-$ and $\bar{K^0}$ mass difference is considered to be large on a 
nuclear physics scale.

The cross section for the kaonic nuclear $1s$ states are shown in table 1.   
The \pik\ reaction for the hypernuclear production shows distinct peaks 
corresponding to series of major shell orbits especially for target nuclei 
with $j_n=\ell_n+1/2$ orbit closed.  We thus take $^{12}$C and $^{28}$Si for 
the present study.  \\

\begin{tabular}{||c|c|c|c||} \hline
nucleus     & $N^{pw}_{eff}$  & $D_{eik}$  & $d\sigma /d \Omega$ mb/sr \\ \hline
$^{12}$C    & 0.055$\sim$0.26   &  0.25 &  0.72$\sim$3.3 \\ 
$^{28}$Si   & 0.029$\sim$0.15   &  0.16 &  0.24$\sim$1.3  \\ \hline
\end{tabular}

\vspace*{0.5cm}
Table 1 \\

Calculated laboratory differential cross sections of the $1s$ states excited 
by the \kmp\ reactions 
at P$_K$=1 GeV/c for the $^{12}$C and $^{28}$Si targets.   Range of values 
corresponds to the $b$ parameter (see text) \\

The calculated cross sections turn out to be quite large which can 
compensate for a low intensity kaon beam.  The large cross section comes 
from the large cross section of the elastic $K + N \rightarrow K + N$ 
reaction and from the transformation of c.m. system to laboratory system.   

Feasibility of the experiment depends on backgrounds.  Dominant backgrounds
are nucleons from knock-out reactions, where kaons are scattered backward 
by the quasifree process.  Since the nucleons associated with the 
deeply-bound kaonic nuclei are much more energetic, the knock-out reactions 
will not be a problem.  

Kaon absorption by two nucleons in nuclei can generate energetic nucleons.  
The process has to involve another nucleon in addition to the \kmN\ 
reaction.   Thus one expects the process gives smaller cross section than 
that of the \kmN\ reaction.   The process can be interpreted as a spreading 
width of the kaonic nuclei.  

A \lam\ produced in the forward direction by the quasifree $(K^-, \pi )$ 
reaction provides an energetic nucleon.   It would not be a serious 
background since its cross section is an order-of-magnitude smaller than 
\kmN\ reaction and no peak structure is expected.

From the experimental point of view energetic protons can be produced by 
knock-out reaction by pions which are contaminated in the kaon beam.   
This process however can be removed by the careful tuning of experimental 
condition.  

It is shown that the \kmp\ and \kmn\ reactions can be used for the study of 
the kaonic nuclei.  Study of the reaction requires intense low energy kaon 
beam for which AGS of BNL and probably PS of KEK are particularly suitable.  
The beam momentum can be chosen by considering the cross section, beam 
intensity and momentum resolution of spectrometer.  There are beam lines 
which provide $K^-$ beam 0.5$\sim$2 GeV/c at BNL and KEK.   The relatively 
broad width ($\sim$10 MeV) and simple structure of the state need 
spectrometers of only modest momentum resolution but wide momentum 
acceptance.   

We demonstrated that the \kmp\ and \kmn\ reaction can be used to obtain 
direct information on the $\bar{K}N$ interaction in nuclear matter.   The 
calculation employed here is rather crude although it is based on 
well-known general concepts in nuclear physics.  

The anther is grateful to discussions with Professors A. Gal, Y. Akaishi, 
T. Tatsumi, and H. Toki.  The anther thanks Dr. R. E. Chrien for careful 
reading of this manuscript.

\bibliographystyle{unsrt}

Figure 1 \\
Diagram for the formation of kaonic nuclei via the $(K^-, N)$ reaction.  
The kaon, the nucleon, and the nucleus are denoted by the dashed, thin 
solid and multiple lines, respectively.  The kaonic nucleus is denoted 
by the multiple lines with the dashed line.  The filled 
circle is the $KN \rightarrow KN$ amplitude while the open circles are the 
nuclear vertices.  The bubbles represent distortion.  \\
\\

\special{epsfile=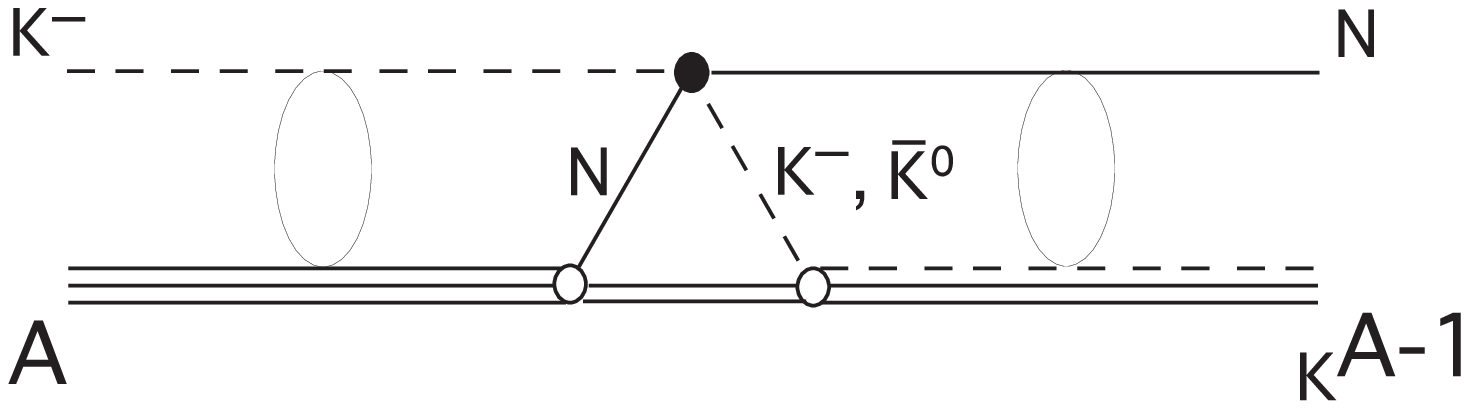 vsize=80}
\vspace*{3.5cm}

Figure 2 \\
The momentum transfer of the $(K^-, N)$ reaction at 0 degrees is shown for 
four reactions.   Here binding energy of kaonic nucleus $^{27}_K{\rm Mg}$ is taken to be -150 MeV.  \\

\special{epsfile=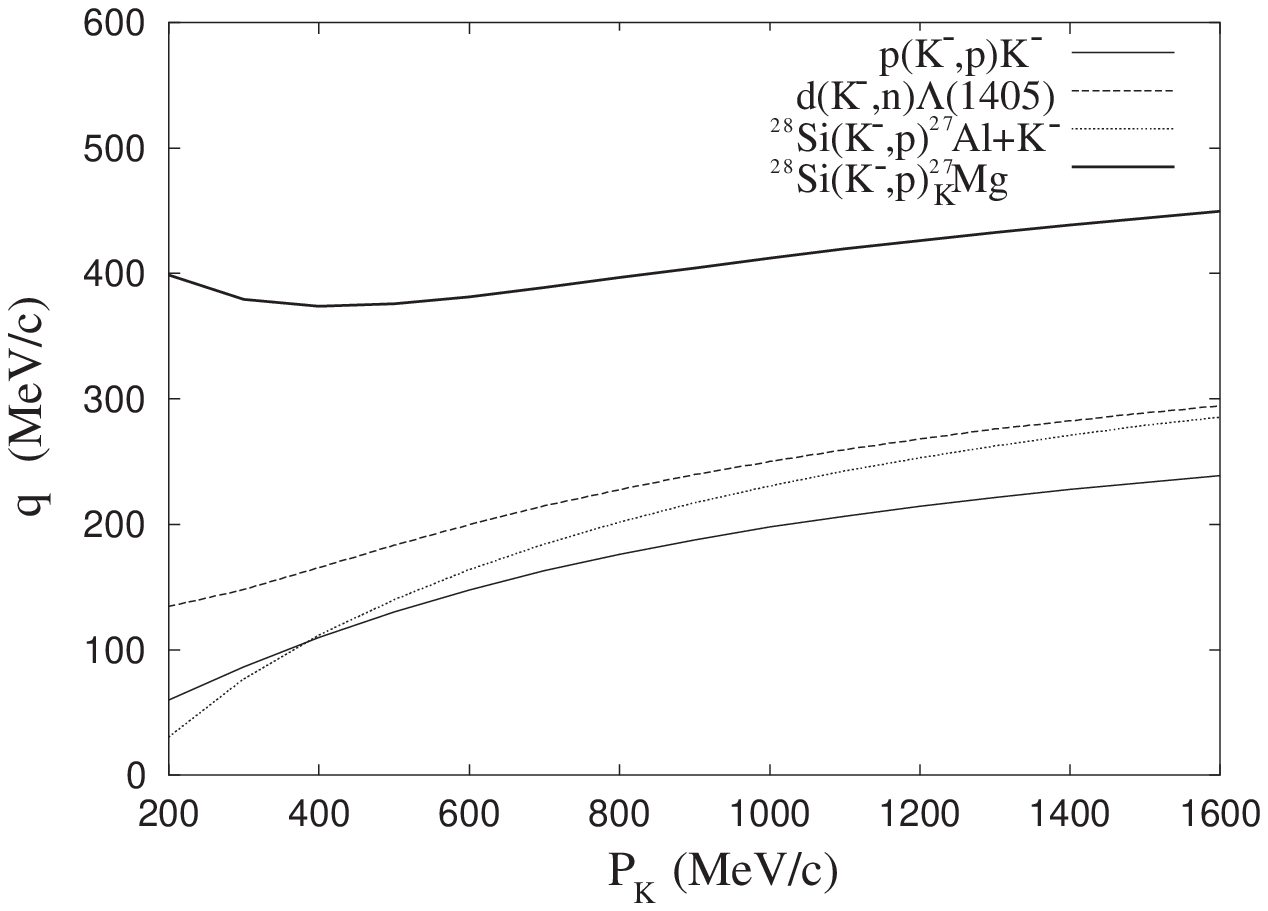 vsize=200}
\vspace*{7.5cm}

Figure 3 \\
The CM differential cross sections of the three reactions are shown as a 
function of incident kaon lab momentum.  \\

\special{epsfile=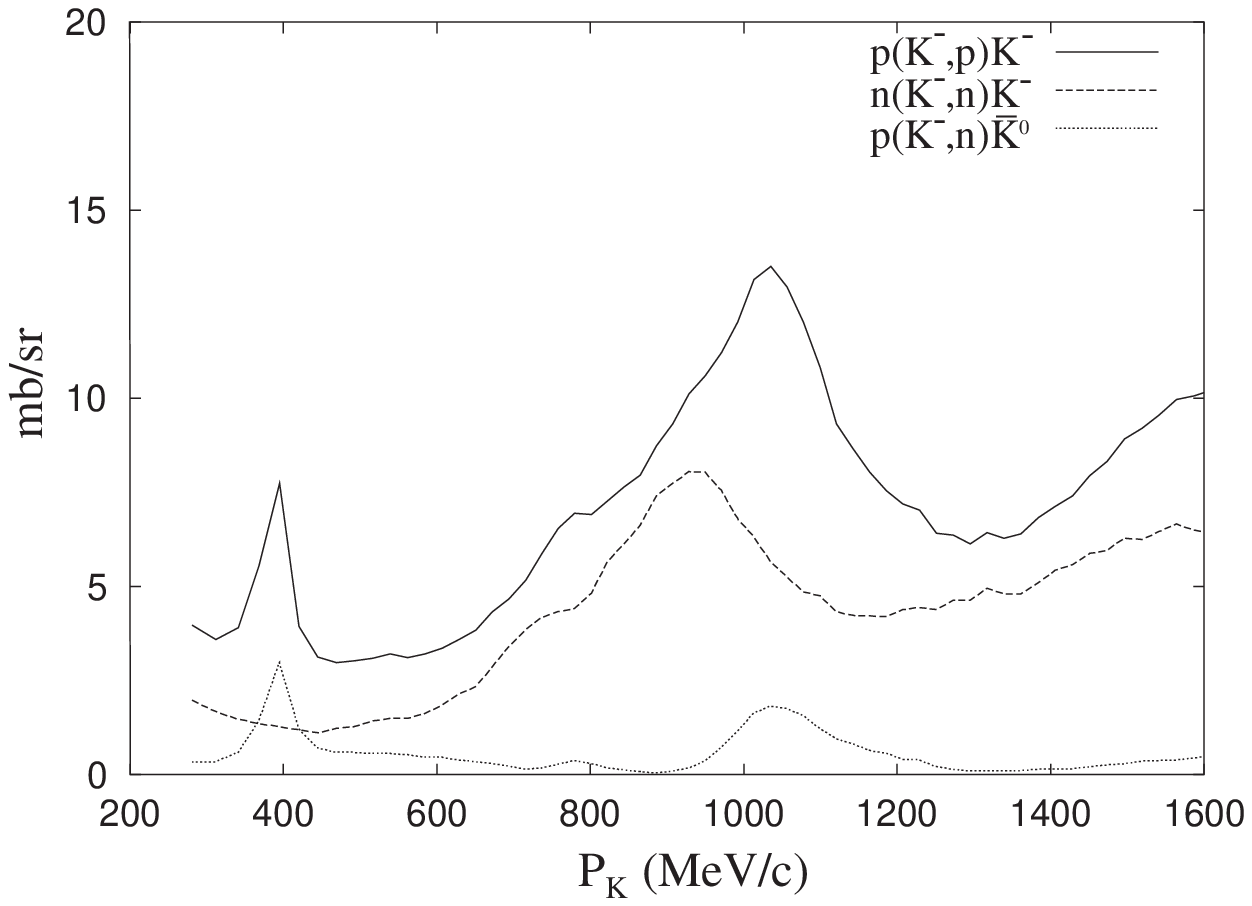 vsize=200}
\vspace*{7.5cm}

\end{document}